\documentclass[preprint,12pt]{elsarticle}

\usepackage{amssymb}
\usepackage{amsmath}
\usepackage{booktabs}
\usepackage{amsthm}
\usepackage{framed} 
\usepackage{multicol} 
\usepackage{tikz}
\usetikzlibrary{arrows.meta, positioning}
\usepackage{xcolor}
\usepackage{graphicx}

\usepackage{graphics}
\usepackage{enumitem}

\usepackage{subcaption}

\usepackage{framed} 
\usepackage{multicol} 

\usepackage{nomencl} 
\makenomenclature
\renewcommand*\nompreamble{\begin{multicols}{2}}
	\renewcommand*\nompostamble{\end{multicols}}
\usepackage[hidelinks]{hyperref}
\begin{document}
	
\makeatletter
\def\ps@pprintTitle{}
\makeatother
\begin{frontmatter}



 \title{PINN vs LSTM: A Comparative Study for Steam Temperature Control in Heat Recovery Steam Generators\tnoteref{label1}}


\author[inst1]{Mojtaba Fanoodi}

\author[inst1]{Farzaneh Abdollahi}

\author[inst2]{Mahdi Aliyari Shoorehdeli}

\affiliation[inst1]{
	organization={Department of Electrical Engineering, AmirKabir University of Technology (Tehran Polythechnique)},
	city={Tehran},
	country={Iran}
}

\affiliation[inst2]{
	organization={Control and Systems Department, Electrical Engineering Faculty, K.N. Toosi University of Technology},
	city={Tehran},
	country={Iran}
}

\begin{abstract}
This paper introduces a direct comparative study of Physics-Informed Neural Networks (PINNs) and Long Short-Term Memory (LSTM) networks for adaptive steam temperature control in Heat Recovery Steam Generators (HRSGs), particularly under valve leakage faults. Maintaining precise steam temperature in HRSGs is critical for efficiency and safety, yet traditional control strategies struggle with nonlinear, fault-induced dynamics. Both architectures are designed to adaptively tune the gains of a PI-plus-feedforward control law in real-time. The LSTM controller, a purely data-driven approach, was trained offline on historical operational data, while the PINN controller integrates fundamental thermodynamic laws directly into its online learning process through a physics-based loss function. Their performance was evaluated using a model validated with data from a combined cycle power plant, under normal load changes and a challenging valve leakage fault scenario. Results demonstrate that while the LSTM controller offers significant improvement over conventional methods, its performance degrades under the unseen fault. The PINN controller consistently delivered superior robustness and performance, achieving a 54\% reduction in integral absolute error compared to the LSTM under fault conditions. This study concludes that embedding physical knowledge into data-driven control is essential for developing reliable, fault-tolerant autonomous control systems in complex industrial applications.
\end{abstract}



\end{frontmatter}

\section{Introduction}
\label{sec1}

\begin{figure}[htbp]
	\centering
	\includegraphics[width=0.8\linewidth]{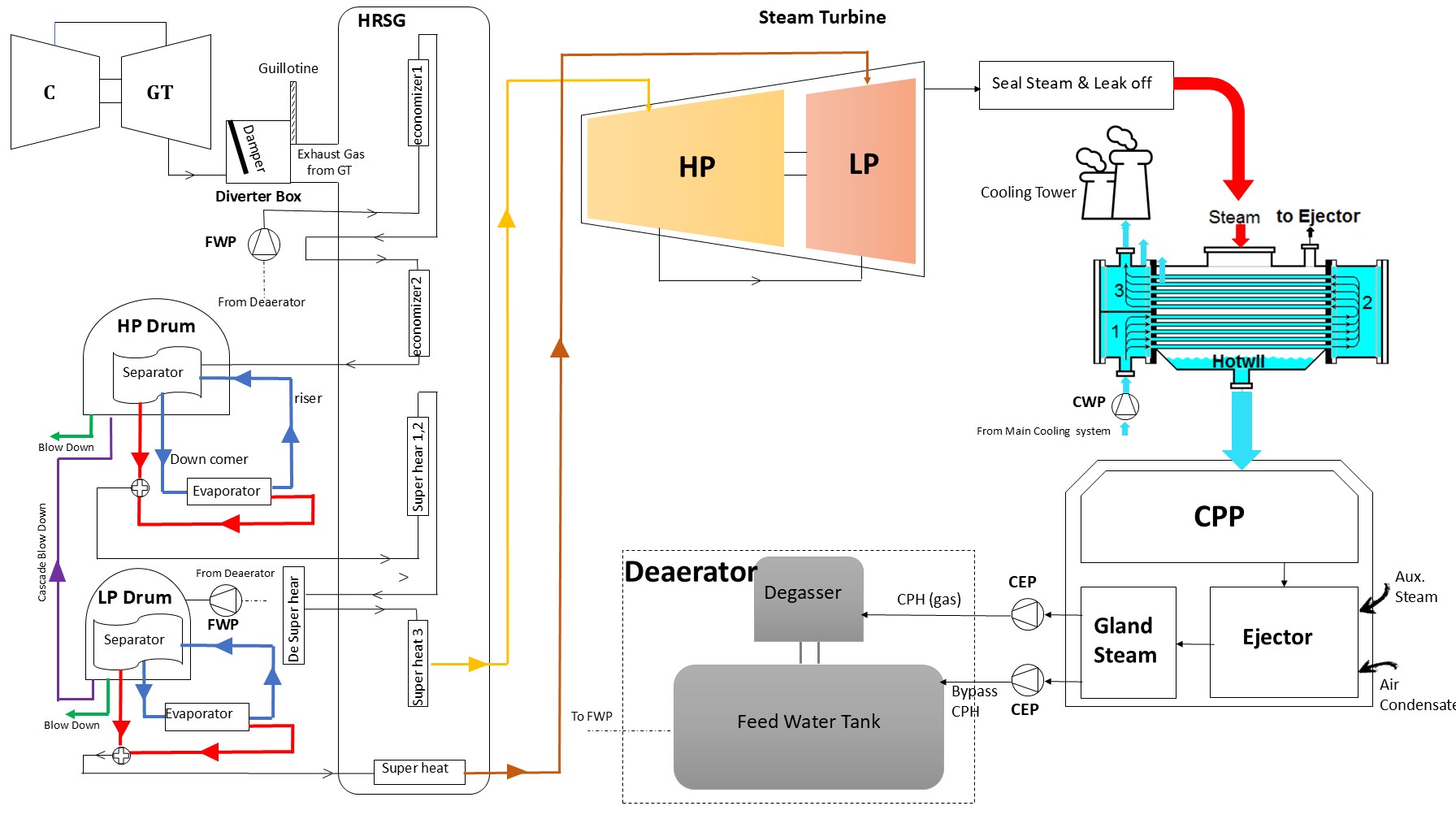}
	\caption{Block diagram of a Combined Cycle Power Plant, showing the gas turbine, HRSG, and steam turbine. The HRSG serves as the critical interface between the gas and steam cycles.}
	\label{fig:ccpp_block}
\end{figure}

Heat Recovery Steam Generators are integral components in combined cycle power plants (CCPP), where they utilize the exhaust heat from gas turbines to generate steam, thereby improving the overall energy efficiency of the system \cite{nadir_thermodynamic_2015,ehyaei2014estimation}. In modern power generation, combined cycle plants account for a large share of installed capacity, and the HRSG serves as the critical interface between the gas turbine and the steam cycle.This overall configuration is illustrated in the block diagram of a typical combined cycle power plant in Fig.~\ref{fig:ccpp_block}.

 A typical HRSG consists of multiple heat-exchange sections, including the economizer, evaporator, superheater, and reheater. Among these, the superheater is particularly sensitive, since it should deliver steam at tightly regulated temperatures to avoid thermal fatigue and efficiency losses in the downstream steam turbine. To regulate this temperature, a desuperheater (or attemperator) injects finely controlled amounts of water into the steam flow. 
 
  The nonlinear evaporation and mixing dynamics of this spray water, combined with transport delays and coupling with upstream and downstream sections, make temperature regulation in the superheater one of the most challenging control problems in HRSG operation. Even small deviations from the setpoint can accelerate material degradation or cause forced outages, underscoring the importance of accurate and robust temperature control. \cite{arpit_state---art_2023, antonova_effect_2017}.

\begin{figure}[htbp]
	\centering
	\includegraphics[width=0.65\linewidth]{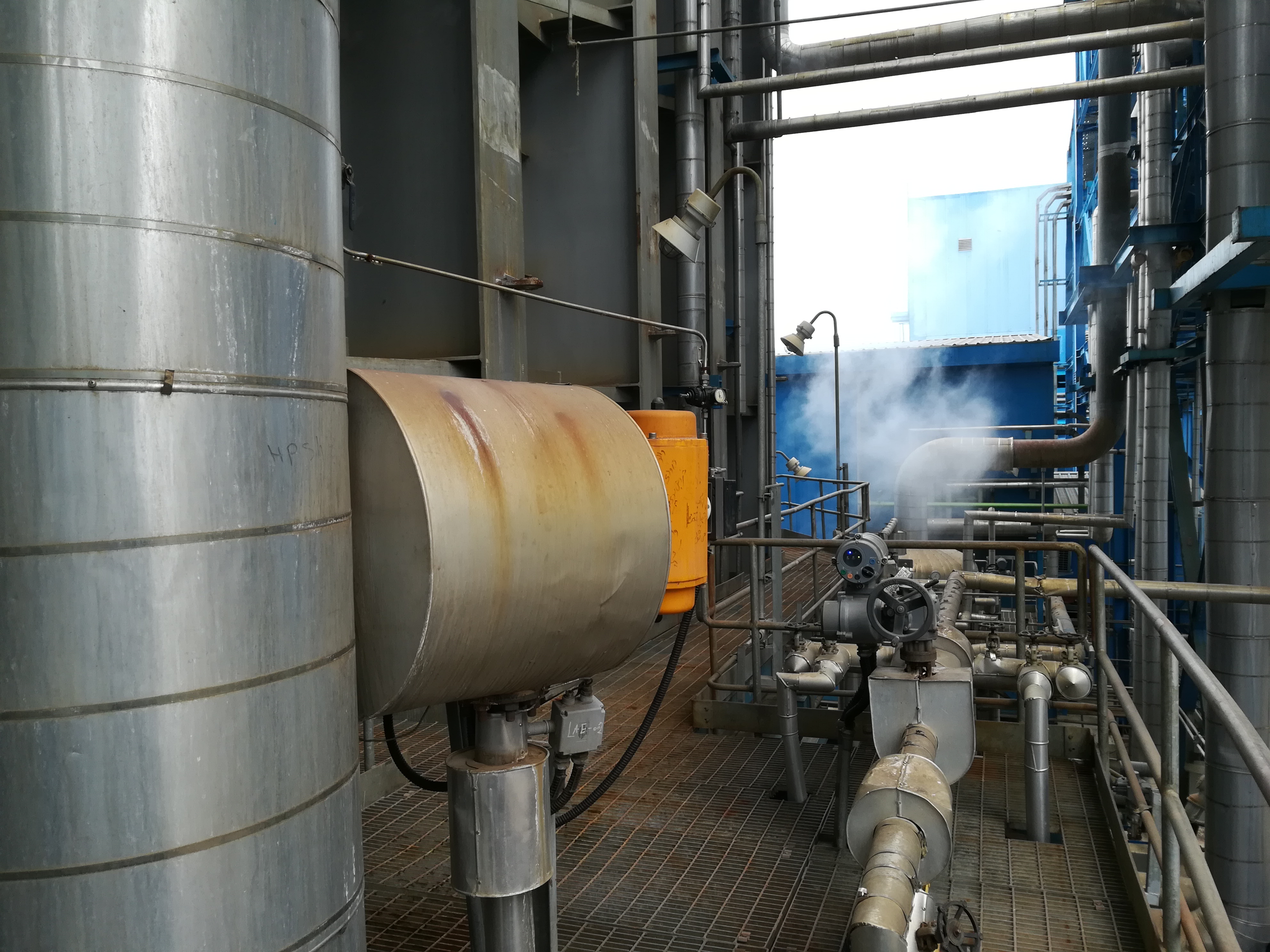}
	\caption{The desuperheater (DSH) unit at the Pareh-Sar power plant, where valve leakage has been identified as a critical fault source affecting steam temperature regulation.}
	\label{fig:dsh_photo}
\end{figure}

In addition to nonlinearities and operational faults such as valve leakage, maintaining stable outlet steam temperature is a challenging control problem. \cite{si2020study,zhao_anomaly_2018}. Temperature fluctuations can cause reduced efficiency, fatigue in heat exchanger tubes, or even emergency shutdowns if limits are exceeded. These challenges are exacerbated by faults such as valve leakage in the desuperheater, often resulting from mechanical wear, erosion, or cracking \cite{zhang_failure_2020, kochmanski_failure_2024}. This leakage introduces uncommanded cooling water into the system, disrupting the thermal balance and leading to significant temperature deviations. For instance, at the Pareh-Sar power plant, an undetected valve leakage fault led to a dangerous temperature overshoot, triggering emergency shutdowns and costly manual interventions\cite{fanoodi2025pinn}. The actual desuperheater unit at the Pareh-Sar power plant is demonstrated in Fig.~\ref{fig:dsh_photo}.

Traditionally, Proportional--Integral--Derivative (PID) controllers have been employed in industry due to their simplicity and ease of implementation. However, fixed-gain PI or PID controllers are inherently reactive and struggle to handle these nonlinear, fault-induced dynamics, variable operating conditions, and persistent disturbances \cite{elhosseini_heat_2022, saha_automatic_2017}. The efforts demonstrated that conventional PI controllers are unable to maintain steam temperature effectively under load variations, motivating the need for more advanced control strategies \cite{li_failure_2022}. Adaptive control, gain scheduling, and model predictive control (MPC) have been proposed as alternatives, offering improved performance at the cost of increased complexity and computation\cite{li_failure_2022,maican_application_2023}.

The limitations of traditional controllers become more pronounced under modern operating conditions. With the growing demand for flexible plant operation, HRSGs should frequently adapt to variable load conditions, start-up/shutdown cycles, and fuel variations. These scenarios expose the shortcomings of fixed-gain controllers and highlight the need for control methods capable of capturing nonlinear dynamics, adapting to unmodeled disturbances, and providing resilience to actuator wear such as valve leakage. Machine learning approaches are good candidates for this task, as they can learn complex temporal relationships directly from plant data without requiring a full first-principles model.

The advent of data-driven methodologies has opened new avenues for advanced process control. Techniques leveraging artificial neural networks (ANNs) have shown remarkable potential in modeling complex, nonlinear systems. Long Short-Term Memory networks, a type of recurrent neural network, are particularly suited for modeling sequential data and temporal dependencies, making them well-suited to capture HRSG dynamics \cite{nelles_nonlinear_2022}. LSTM-based methods have been applied for predictive control of steam temperature, \cite{zhu2023modelling} showing that an LSTM-based model could capture nonlinear coupling effects and outperform PID controllers. Hybrid frameworks have been further extended through the proposal of an LSTM–Particle Swarm Optimization (PSO) strategy and the introduction of a Dense Residual LSTM-Attention Network, which reduced prediction error and enabled early warnings for over-temperature conditions. The potential of LSTM architectures for predictive and corrective steam temperature control is thereby highlighted \cite{jin2023lstm, tong2022dense}.

Nevertheless, purely data-driven methods such as LSTM require large volumes of data for training and often lack robustness when operating conditions deviate from the training dataset or when faults occur. This motivates physics-guided machine learning approaches. Physics-Informed Neural Networks embed governing equations and thermodynamic principles directly into the training process, combining data-driven flexibility with physical consistency \cite{raissi_physics-informed_2019, bolderman2024physics}. By embedding known thermodynamic equations directly into the learning process as a regularization term, PINNs ensure that their predictions and control actions are not only accurate but also adhere to physical constraints. This hybrid approach promises greater robustness, interpretability, and performance, especially in fault conditions that may not be fully represented in historical datasets. In HRSGs, PINNs have been shown to enhance steam temperature regulation by reducing overshoot, improving adaptability, and increasing fault tolerance in valve leakage . Previous works successfully developed a fault-tolerant control framework for an HRSG superheater using a PINN to dynamically adapt controller gains, demonstrating significant performance improvements over conventional methods under valve leakage faults\cite{fanoodi2025pinn}.

From an industrial perspective, a direct comparative study is crucial. Plant operators and engineers require not only high accuracy in prediction but also demonstrable robustness under practical fault scenarios. While LSTM-based strategies may deliver strong performance in well-trained regimes, their ability to generalize to unseen disturbances remains uncertain. On the other hand, PINNs introduce physical consistency. However they have not been systematically benchmarked against purely data-driven solutions in HRSG applications. A side-by-side comparison therefore provides valuable insights for guiding the adoption of learning-based controllers in safety-critical power plant environments.

The rest of this paper is structured as follows: Section 2 describes the HRSG system and formulates the control problem. Section 3 details the development of the LSTM and PINN-based adaptive controllers. Section 4 presents the simulation setup and results, comparing the performance of all controllers. Finally, Section 5 provides the concluding remarks.

\section{System Description and Problem Formulation}

This section describes the key components of the HRSG system under study and formulates the steam temperature control problem, including the challenge posed by valve leakage faults.

\subsection{HRSG System Overview}
\begin{figure}[htbp]
	\centering
	\includegraphics[width=0.95\linewidth]{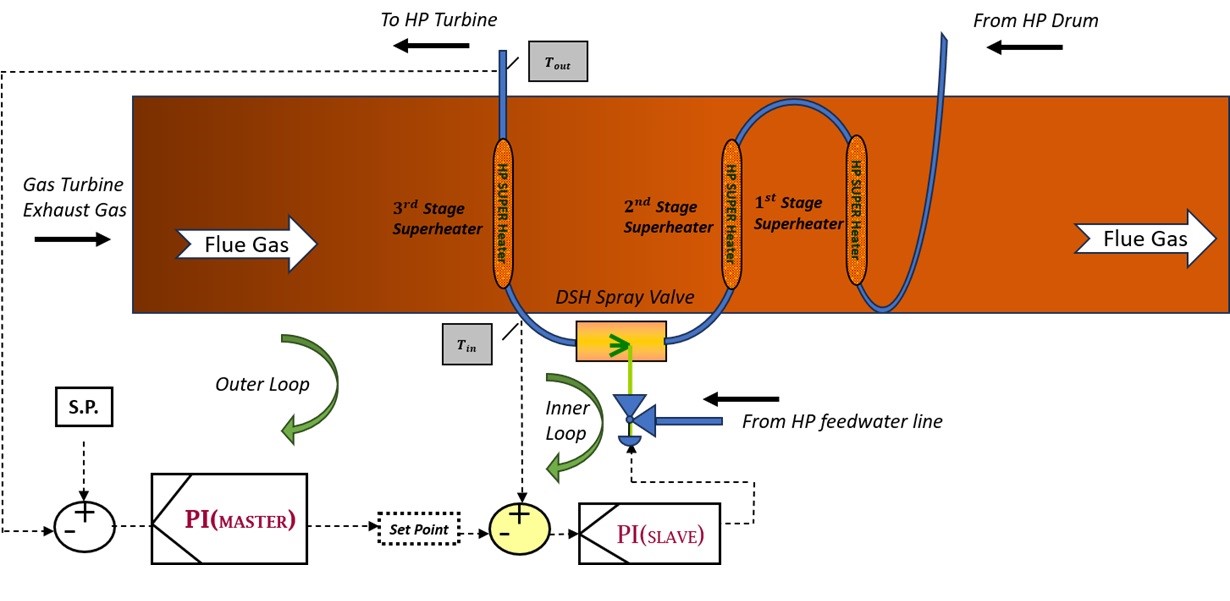}
	\caption{Schematic of the desuperheater process. Spray water is injected into the superheated steam flow, and evaporation/mixing dynamics regulate outlet steam temperature.}
	\label{fig:dsh_schematic}
\end{figure}

The focus of this study is on the superheater and desuperheater sections of a dual-pressure HRSG, a critical part of a combined cycle power plant. The primary function of this system is to deliver superheated steam at a precisely controlled temperature to the steam turbine. The internal spray-water injection and mixing process inside the desuperheater is schematically shown in Fig.~\ref{fig:dsh_schematic}.

The process can be summarized as follows:
\begin{itemize}
	\item \textbf{Superheater (SH):} Exhaust gases from the gas turbine transfer heat to saturated steam, raising its temperature significantly above the saturation point.
	\item \textbf{Desuperheater (DSH):} To prevent overheating and maintain the final steam temperature at its desired setpoint, a control valve sprays cooling water into the steam flow.
\end{itemize}

The main manipulated variable is the spray water flow rate, denoted as $u(t)$ [kg/s]. The primary controlled output is the steam temperature at the outlet of the desuperheater, denoted as $T_{\text{out}}^{\text{(dsh)}}(t)$ or $y(t)$ [°C]. The most significant measurable disturbance is the gas turbine exhaust temperature, $T_{\text{gt}}(t)$ [°C], as it directly affects the heat input to the superheater.

\subsection{Dynamic Model}
Based on first principles of mass and energy balance \cite{chaibakhsh_modelling_2013, maffezzoni_boiler-turbine_1997}, the dynamics of the superheater and desuperheater can be captured by a set of nonlinear differential equations. For the purpose of this control-oriented study, the model can be conceptually represented in a state-space form.

Let the system states be defined as:
\begin{align*}
	x_1(t) &= T_{\text{out}}^{\text{(sh)}}(t) \quad \text{(Outlet temperature of the superheater)} \\
	x_2(t) &= T_{\text{out}}^{\text{(dsh)}}(t) \quad \text{(Outlet temperature of the desuperheater)}
\end{align*}

The system dynamics are then given by\cite{fanoodi2025pinn}:

\begin{equation}
	\begin{cases}
		\dot{x}_1(t) = K_2(K_1 d_1 + d_7(x_2(t) - x_1(t)) - d_3)\\
		\dot{x}_2(t) = K_3\Big[(d_2 + u(t))(d_4 - x_2(t)) - u(t)(d_4 - d_5) + \bar{m}_{in_{dsh}} d_6\Big]  \\
		y(t) = x_1(t)
	\end{cases} \label{nof}
\end{equation}

where $u(t)$ is the control input (spray water flow rate). The disturbances $d_i$ encompass key operational variables:
\begin{align*}
	&d_1 = \dot{m}_{\text{fuel}}, \quad d_2 = \dot{m}_{\text{steam}}, \quad d_3 = \dot{T}_a, \\
	&d_4 = T_{\text{in}}^{\text{(dsh)}}, \quad d_5 = T_{\text{spray}}, \quad d_6 = \dot{T}_{\text{in}}^{\text{(dsh)}}, \quad d_7 = \dot{m}_{\text{out}}^{\text{(dsh)}}
\end{align*}

The state equations are derived from thermodynamic principles and are inherently nonlinear, capturing the complex heat transfer and mixing phenomena within the HRSG.

\subsection{Valve Leakage Fault Model}
\begin{figure}[h!]
	\centering
	\includegraphics[width=0.95\linewidth]{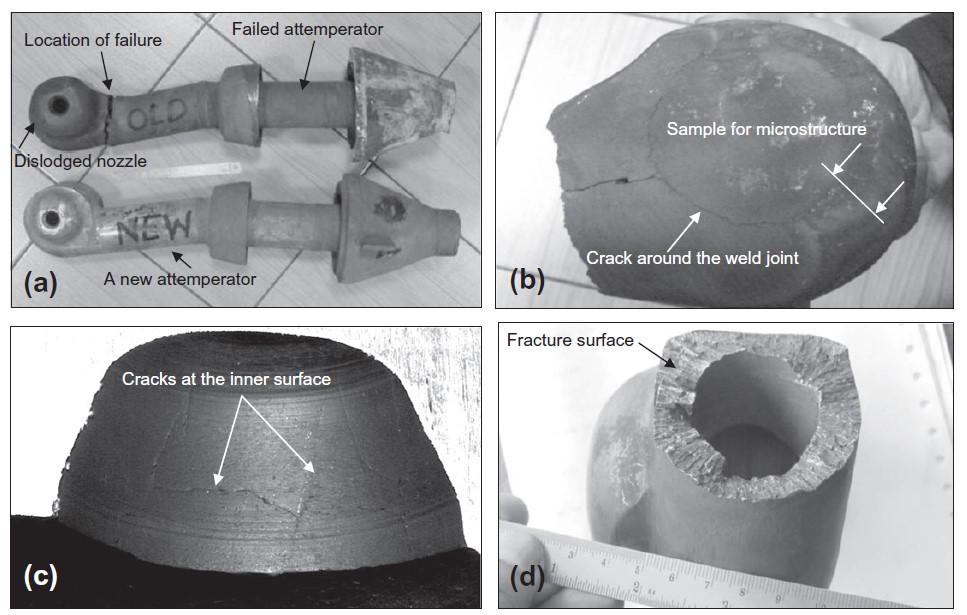}
	\caption{Visual examination of the damaged attemperator: (a) displacement of the spray nozzle tip, (b) surface cracks near the weld joint, (c) multiple internal cracks along the nozzle inner surface, and (d) fractured surface exhibiting an irregular, fibrous texture.\cite{mukhopadhyay_failure_2011}}
	\label{fault}
\end{figure}

A common and critical fault in this system is leakage in the spray water control valve. This fault manifests as an uncommanded, persistent flow of cooling water, even when the valve is instructed to be fully closed ($u(t) = 0$). Figure \ref{fault} illustrates a sample of such faults.

To model this scenario, the control input in the desuperheater dynamics can expressed:
\begin{equation}
\dot{x}_2 = K_3\left[(d_2 + u + f)(d_4 - x_2) - (u + f)(d_4 - d_5) + \bar{m}_{in_{dsh}} d_6\right]
\end{equation}

where $f \geq 0$ represents the constant leakage flow rate [kg/s], bounded by a maximum value $f_{\text{max}}$. This fault term $f$ continuously perturbs the system, disrupting the energy balance and making temperature regulation with a standard controller significantly more challenging.

\subsection{Control Objective}
The primary control objective is to maintain the outlet steam temperature $y(t)$ at a desired setpoint $r(t)$, considering:
\begin{itemize}
	\item Variations in the gas turbine exhaust temperature $T_{\text{gt}}(t)$ (major disturbance).
	\item Changes in steam mass flow rate due to load changes.
	\item The presence of a valve leakage fault $f$.
\end{itemize}

This is achieved by designing a controller that computes the necessary spray water flow rate $u(t)$.

\subsection{Control Strategy}
To achieve this objective, a PI-plus-feedforward control law is proposed:
\begin{equation}
	u(t) = K_p(t) \cdot e(t) + K_i(t) \int_0^t e(\tau) d\tau + K_{ff}(t) \cdot T_{\text{gt}}(t)
	\label{eq:control_law}
\end{equation}
where $e(t) = r(t) - y(t)$ is the temperature tracking error.

This PI+feedforward formulation is consistent with standard industrial practice in HRSGs, where the \textit{feedforward} term $K_{ff}(t) \cdot T_{\text{gt}}(t)$ is crucial for anticipating the effect of changes in exhaust gas temperature, providing proactive compensation. The \textit{feedback} PI terms correct for any remaining error.

The core challenge is the real-time adjustment of the gains $K_p(t)$, $K_i(t)$, and $K_{ff}(t)$ to maintain optimal performance across varying loads and under the fault condition $f$. This paper compares two distinct neural network-based approaches to solve this gain tuning problem:
\begin{itemize}
	\item A purely \textbf{data-driven} method using an LSTM network.
	\item A \textbf{physics-guided} method using a PINN.
\end{itemize}
Section~\ref{sec:methodology} presents the architectures and training procedures of the proposed LSTM and PINN-based adaptive controllers.

\section{Methodology: Data-Driven Controllers}
\label{sec:methodology}

This section details the architecture, training, and implementation of the two data-driven controllers developed for this study: the Long Short-Term Memory network and the Physics-Informed Neural Network. Both are designed to adaptively tune the gains \( K_p(t) \), \( K_i(t) \), and \( K_{ff}(t) \) of the PI-plus-feedforward control law (Eq. \ref{eq:control_law}) in real-time.

\subsection{LSTM-Based Adaptive Controller}
The LSTM-based controller is a purely data-driven approach that learns the optimal mapping from the system recent state history to the control gains.

\subsubsection{Network Inputs and Outputs}
The LSTM network is designed to capture temporal dependencies in the system dynamics. Its inputs at time \( t \) are a sliding window of the \( N \) most recent time steps of the following relevant signals:
\begin{equation}
	\mathbf{I}_{\text{LSTM}}(t) = [ \mathbf{s}(t), \mathbf{s}(t-1), ..., \mathbf{s}(t-N+1) ]^T
\end{equation}
where the state vector \( \mathbf{s}(t) \) is:
\begin{equation}
	\mathbf{s}(t) = [ e(t), y(t), T_{\text{gt}}(t), u(t) ]
\end{equation}
and  $ T_{\text{gt}}(t) $ is the gas turbine exhaust temperature (measured disturbance).

The output of the LSTM network is the vector of adaptive gains:
\begin{equation}
	\mathbf{O}_{\text{LSTM}}(t) = [ K_p(t), K_i(t), K_{ff}(t) ]
\end{equation}

The dataset used for training and evaluation was based on a hybrid approach. 
First, a total of 120 hours (five days) of operating data were generated 
using the validated nonlinear HRSG model described in Section~2, calibrated with 
measurements from Unit~1 of the Pareh-Sar power plant. This simulated dataset, 
covering normal load ramps, setpoint changes, and variations in steam mass flow 
rate, was used to pretrain the LSTM network. The model was then fine-tuned on 
real operational records from Pareh-Sar to capture plant-specific noise and dynamics. 
Both simulated and real datasets were divided into 70\% for training, 15\% for 
validation, and 15\% for testing. Importantly, only fault-free operation was 
included in the training and validation sets, while the valve leakage fault 
scenario was reserved exclusively for the testing stage. This ensured that the 
LSTM network learned from representative but fault-free dynamics, allowing its 
generalization capability to be fairly assessed under the unseen fault condition.

\subsubsection{Network Architecture and Training}
The proposed LSTM architecture consists of:
\begin{itemize}
	\item An input layer that accepts the sequence \( \mathbf{I}_{\text{LSTM}}(t) \).
	\item Two LSTM layers with 50 and 25 units, respectively, using hyperbolic tangent (tanh) activation functions to capture complex temporal patterns.
	\item A dropout layer after each LSTM layer to prevent overfitting.
	\item A final dense output layer with 3 units and linear activation to generate the gain values.
\end{itemize}

The network was trained \textit{offline} on a historical dataset \( \mathcal{D} = \{ (\mathbf{I}^{(i)}, \mathbf{K}^{* (i)}) \} \) from the power plant, encompassing normal operation and various disturbance scenarios. The target gains \( \mathbf{K}^{*} = [K_p^*, K_i^*, K_{ff}^*] \) were not directly measurable; instead, the dataset was generated by simulating the system model and using an optimization algorithm  to find the ideal gains that minimized the integral of squared error (ISE) of the temperature tracking for each sequence of states in the historical data.
The network was trained to minimize the Mean Squared Error (MSE) between its predictions and these optimal target gains:
\begin{equation}
	L_{\text{LSTM}} = \frac{1}{M} \sum_{i=1}^{M} \left\| \mathbf{O}_{\text{LSTM}}^{(i)} - \mathbf{K}^{* (i)} \right\|^2
\end{equation}
The Adam optimizer was used with a learning rate of 0.001.

\subsubsection{Implementation}
\begin{figure}[htbp]
	\centering
	\includegraphics[width=0.95\linewidth]{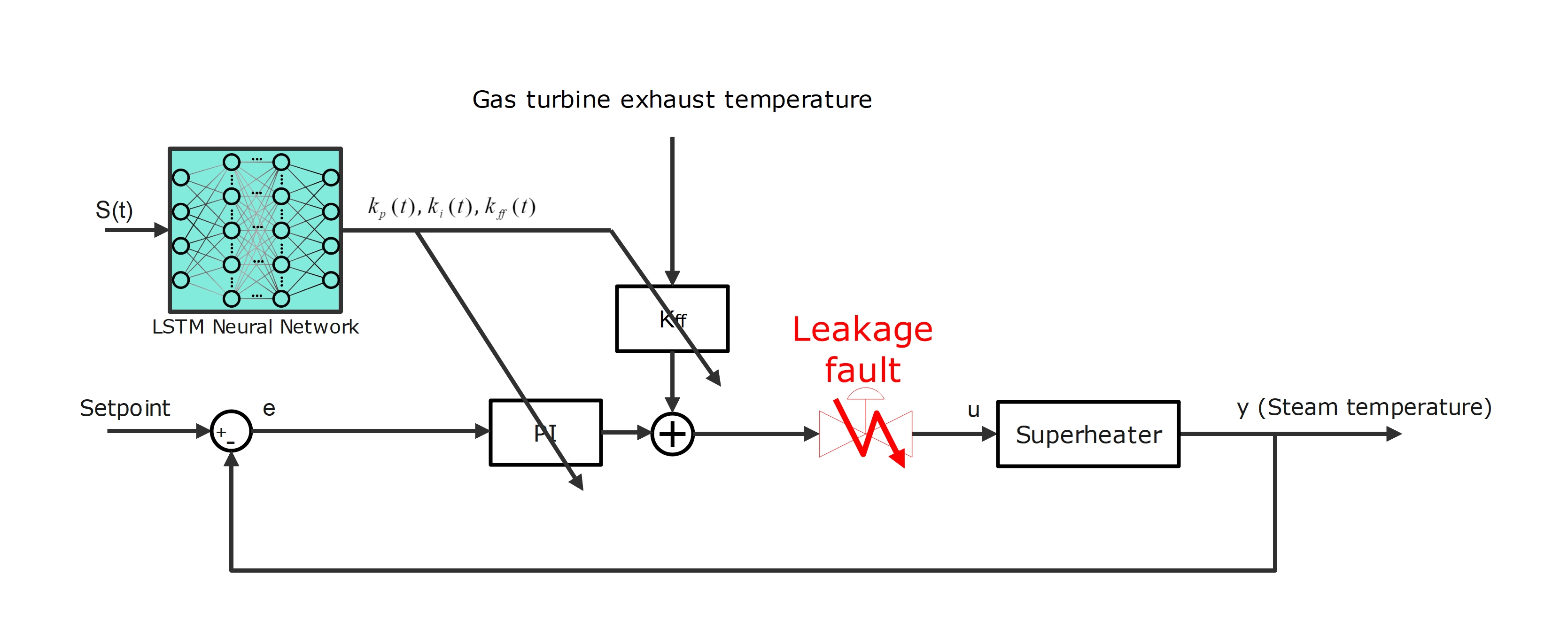}
	\caption{Control loop structure illustrating the LSTM-based gain tuning strategy for HRSG superheater temperature regulation, with the leakage fault modeled in the spray-water valve actuation.}
		
	\label{fig:lstm_fault_loop}
\end{figure}

Once trained, the LSTM network was deployed as an adaptive gain tuner, as shown in Figure~\ref{fig:lstm_fault_loop}. At each control interval, the past \( N \) samples of the input states are fed into the network, which then outputs the three gains for use in the control law (Eq. \ref{eq:control_law}) under a leakage fault in the spray-water valve actuation.

\subsection{PINN-Based Adaptive Controller}
The PINN-based controller follows a similar adaptive gain-tuning structure. It incorporates the known physics of the HRSG system directly into its learning process, enabling more robust online adaptation.

\subsubsection{Network Inputs and Outputs}
In contrast to the LSTM, the PINN is a feedforward network that operates on the \textit{current} system state, as the physical laws are memoryless. Its inputs are:
\begin{equation}
	\mathbf{I}_{\text{PINN}}(t) = [ e(t), y(t), T_{\text{gt}}(t) ]
\end{equation}
The output is the same gain vector:
\begin{equation}
	\mathbf{O}_{\text{PINN}}(t) = [ K_p(t), K_i(t), K_{ff}(t) ]
\end{equation}

\subsubsection{Network Architecture and Physics-Informed Loss}
The PINN architecture consists of:
\begin{itemize}
	\item An input layer with 3 neurons.
	\item Four fully connected hidden layers with 256, 128, 64, and 32 neurons, respectively, using tanh activation.
	\item An output layer with 3 linear neurons.
\end{itemize}

The key innovation is the custom \textit{physics-informed loss function} used to train the network online. The total loss \( L_{\text{total}} \) is a weighted sum of a data loss term and a physics loss term:
\begin{equation}
	L_{\text{total}} = L_{\text{data}} + \mu L_{\text{physics}}
	\label{eq:pin_loss}
\end{equation}
where \( \mu \) is a weighting hyperparameter.

The \textbf{data loss} \( L_{\text{data}} \) is the MSE of the tracking error, ensuring the controller minimizes the primary objective:
\begin{equation}
	L_{\text{data}} = \frac{1}{2} e(t)^2
\end{equation}

The \textbf{physics loss} \( L_{\text{physics}} \) ensures that the gains produced by the network lead to control actions that respect the underlying system dynamics. It penalizes the residual of the state-space model:
\begin{equation}
	L_{\text{physics}} = \frac{1}{2} \left( \left\| \dot{x}_1 - f_1(\cdot) \right\|^2 + \left\| \dot{x}_2 - f_2(\cdot) \right\|^2 \right)
\end{equation}

where $f_1(\cdot)$ and $f_2(\cdot)$ denote the right-hand sides of the superheater and desuperheater dynamics as defined in Eq.~\ref{nof}.
The derivatives \( \dot{x}_1, \dot{x}_2 \) are computed using automatic differentiation through the network, and the functions \( f_1, f_2 \) are evaluated using the current states and the control action \( u(t) \) generated by the PINN-based gains.

Unlike the LSTM controller, the PINN does not rely on a large pre-generated dataset or an explicit train/test split. Instead, it is trained online in real time, continuously updating its parameters based on the current operating data and the embedded physics-informed loss. This design eliminates the need for fault-inclusive training data, since the physics constraints act as a built-in regularizer that guides learning even under unseen conditions such as valve leakage. Consequently, the PINN can adapt to novel disturbances directly during operation, without requiring prior exposure to similar fault scenarios.

\subsubsection{Online Training and Implementation}
\begin{figure}[htbp]
	\centering
	\includegraphics[width=0.95\linewidth]{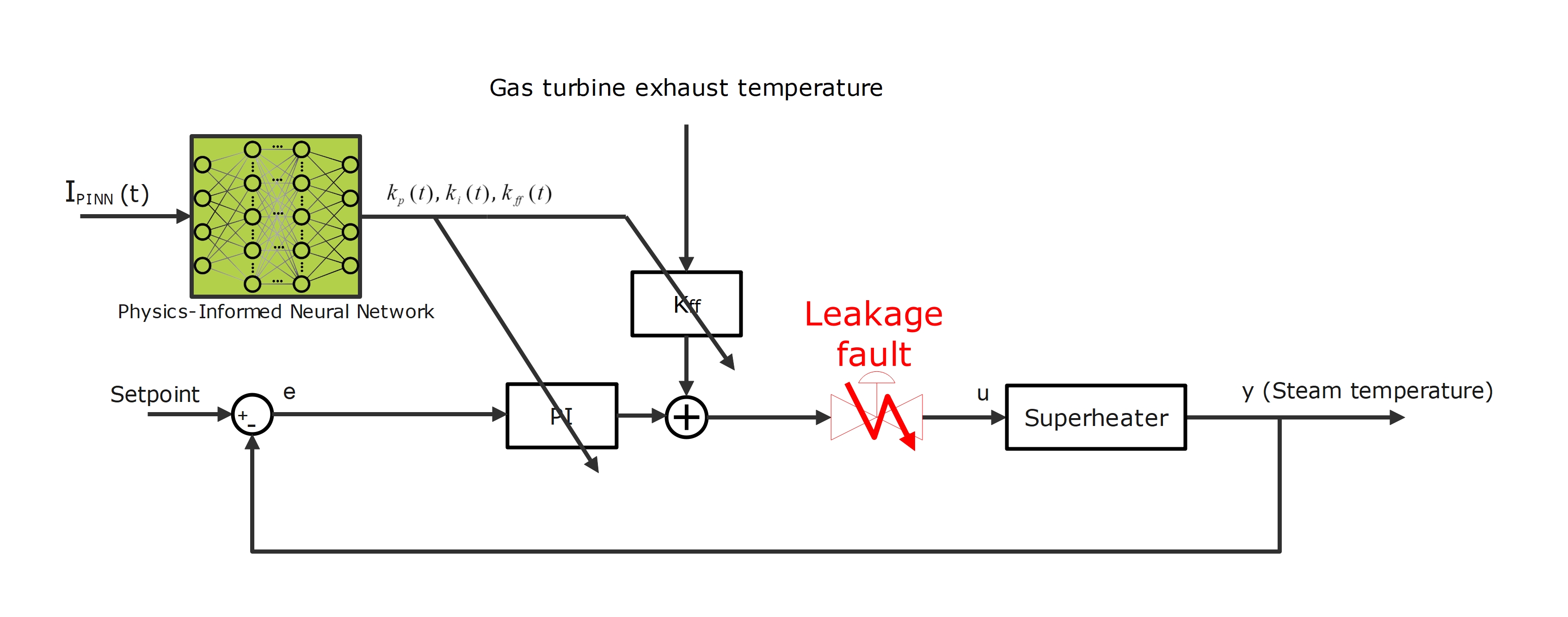}
	\caption{Control loop structure where a PINN is employed to tune the controller gains for HRSG superheater temperature regulation under a leakage fault in the spray-water valve actuation.}
	\label{fig:pinn_fault_loop}
\end{figure}

The PINN is trained \textit{online} using a gradient descent approach. The network parameters (weights and biases) are continuously updated at each time step \( t \) to minimize the total loss \( L_{\text{total}} \):
\begin{equation}
	\theta(t+1) = \theta(t) - \eta \nabla_{\theta} L_{\text{total}}
\end{equation}
where \( \theta \) are the network parameters and \( \eta \) is the learning rate. This process allows the PINN to continuously adapt its gain-tuning strategy based on real-time performance (data loss) while being guided by the immutable laws of physics (physics loss), making it inherently robust to novel faults like valve leakage. Fig~\ref{fig:pinn_fault_loop} illustrates the control loop structure where the PINN tunes the controller gains for HRSG superheater temperature regulation under a leakage fault in the spray-water valve actuation.

Table \ref{tab:method_comparison} summarizes the key characteristics of the two developed adaptive controllers.

\begin{table}[h!]
	\centering
	\caption{Comparison of LSTM and PINN Controller Properties}
	\label{tab:method_comparison}
	\begin{tabular}{p{0.22\textwidth}p{0.33\textwidth}p{0.33\textwidth}}
		\hline
		\textbf{Property} & \textbf{LSTM-Based Controller} & \textbf{PINN-Based Controller} \\ \hline
		\textbf{Core Principle} & Purely data-driven, learns temporal patterns from historical data. & Hybrid; combines data-driven learning with physical laws. \\ \hline
		\textbf{Training} & Offline, supervised learning on historical data. & Online, continuous learning via gradient descent. \\ \hline
		\textbf{Inputs} & Sequence of past states (\( e, y, T_{gt}, u \)). & Current state only (\( e, y, T_{gt} \)). \\ \hline
		\textbf{Physics Incorporation} & None. Relies entirely on data correlations. & Explicitly embedded via physics-based loss term. \\ \hline
		\textbf{Primary Strength} & Improved tracking under known conditions; reduced overshoot ($\approx 6\,^{\circ}\mathrm{C}$) and IAE ($685.7\,^{\circ}\mathrm{C\cdot s}$) compared to PI. & High robustness under unseen faults; smallest overshoot ($\approx 1.8\,^{\circ}\mathrm{C}$) and lowest IAE ($314.8\,^{\circ}\mathrm{C\cdot s}$). \\ \hline
		\textbf{Primary Weakness} & Performance degrades under conditions not seen during training. Residual offset remained in leakage case. & Requires definition of governing equations. Moderate computational overhead due to online gradient updates. \\ \hline
	\end{tabular}
\end{table}

\section{Simulation Setup and Results}
\label{sec:results}

This section presents the simulation environment, operational scenarios, and a detailed comparative analysis of the control strategies under normal and faulty conditions.

\subsection{Simulation Setup and Parameters}
To validate the proposed controllers, a nonlinear model of the HRSG superheater and desuperheater, based on the equations in Section 2, was implemented in Python. The model parameters were calibrated and validated using real operational data from Unit 1 of the Pareh-Sar Combined Cycle Power Plant, ensuring the simulations accurately reflect true plant dynamics.

The controllers were evaluated under identical conditions. The nominal operating parameters used in the simulation are summarized in Table \ref{tab:sim_params}.

\begin{table}[h!]
	\centering
	\caption{Simulation Parameters and Nominal Operating Conditions}
	\label{tab:sim_params}
	\begin{tabular}{lc}
		\hline
		\textbf{Parameter} & \textbf{Value} \\ \hline
		Nominal Steam Mass Flow Rate, $\dot{m}_{\text{steam}}$ & 65 kg/s \\
		Nominal Setpoint, $r(t)$ & 515 $^\circ$C \\
		Gas Turbine Exhaust Temp., $T_{\text{gt}}(t)$ & 530--580 $^\circ$C \\
		Nominal Spray Water Temp., $T_{\text{spray}}$ & 150 $^\circ$C \\
		Control Input Constraint, $u(t)$ & [0, 2] kg/s \\
		Simulation Time Step & 1 second \\
		\hline
	\end{tabular}
\end{table}

Three controllers were compared:
\begin{itemize}
	\item \textbf{ PI:} A conventional fixed-gain PI controller with manually tuned gains ($K_p=1.2$, $K_i=105$), representing current industry practice.
	\item \textbf{LSTM Controller:} The data-driven gain scheduler described in Section 3.1.
	\item \textbf{PINN Controller:} The physics-informed adaptive controller described in Section 3.2.
\end{itemize}

\subsection{Operational and Fault Scenario}
\begin{figure}[h!]
	\centering
	\begin{subfigure}{0.9\textwidth}
		\centering
		\includegraphics[width=0.75\linewidth]{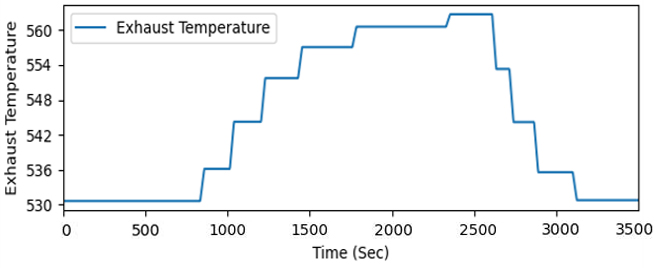}
		\caption{Changes of temperature in gas turbine exhaust.}
	\end{subfigure}
	
	\vspace{0.5cm} 
	
	\begin{subfigure}{0.9\textwidth}
		\centering
		\includegraphics[width=0.75\linewidth]{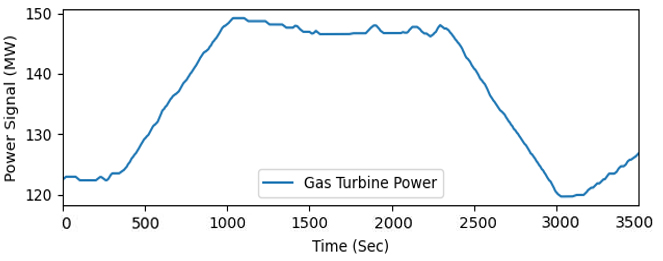}
		\caption{Load changes in MW.}
	\end{subfigure}
	
	\caption{Operational scenario showing (a) changes of temperature in the gas turbine exhaust and (b) corresponding load changes in MW.}
	\label{fig:load_temp_changes}
\end{figure}

The controllers were evaluated under a combined operational and fault condition to assess their robustness. In this scenario, the gas turbine power output was increased at a rate of 6 MW/min, resulting in a ramp increase of the exhaust gas temperature $T_{gt}(t)$ from 530$^\circ$C to 560$^\circ$C. Simultaneously, an unmeasured valve leakage fault was presented. This setup imposes both a large predictable disturbance and an actuator fault, providing a stringent test of the controller fault-tolerant capability. Fig \ref{fig:load_temp_changes} shows the variations in the gas turbine exhaust temperature together with the corresponding load changes in MW.

\subsection{Performance Metrics}
The following key performance indicators (KPIs) were used for quantitative comparison:
\begin{itemize}
	\item \textbf{Integral of Absolute Error (IAE):} $\text{IAE} = \int |e(t)| dt$. Measures total tracking error.
	\item \textbf{Maximum Overshoot (MO):} $\text{MO} = \max(0, y(t) - r(t))$. The highest peak above the setpoint.
	\item \textbf{Settling Time ($T_s$):} The time required for the output $y(t)$ to enter and remain within a $\pm 1^\circ$C band of the setpoint after a disturbance.
	\item \textbf{Control Effort Variance (CEV):} $\text{Var}(u(t))$. Measures the smoothness of the control signal; lower variance indicates smoother actuator movement and less wear.
\end{itemize}

\subsection{Results and Discussion}

\subsubsection{Response to Load Change with Valve Leakage Fault}
\begin{figure}[h!]
	\centering
	\includegraphics[width=0.95\linewidth]{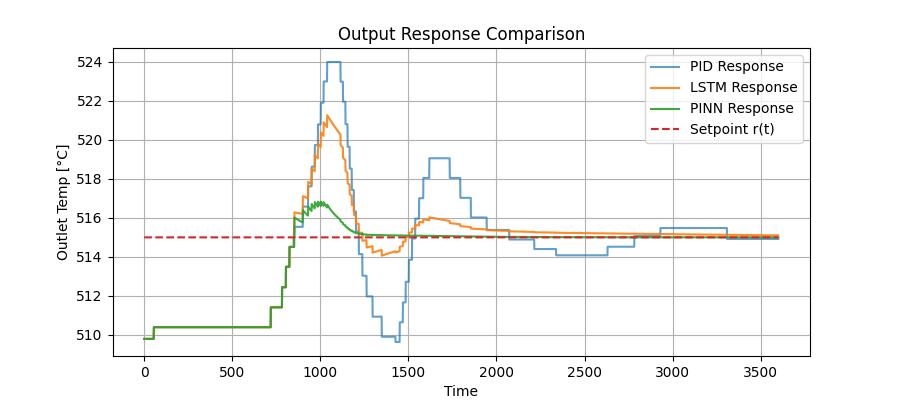}
	\caption{Outlet temperature response under the combined load ramp and valve leakage fault.}
	\label{fig:scenario2_temp}
\end{figure}

\begin{figure}[h!]
	\centering
	\includegraphics[width=0.95\linewidth]{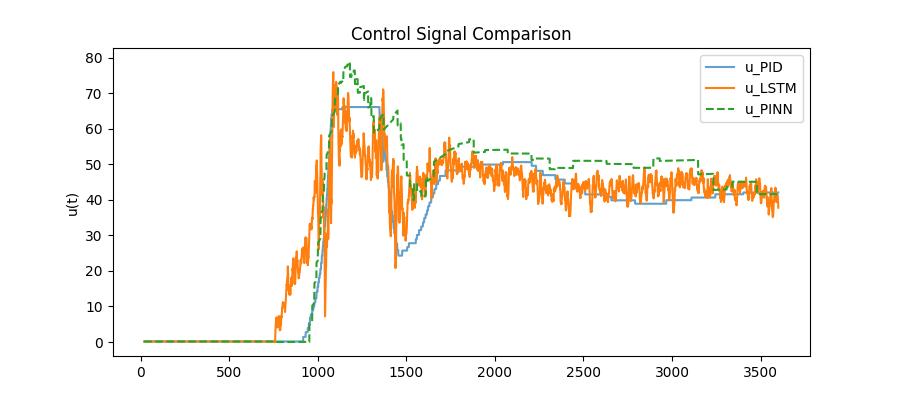}
	\caption{Control input for the three controllers during the combined scenario. Note the PINN increased control action to compensate for the leakage.}
	\label{fig:scenario2_control}
\end{figure}

Figures \ref{fig:scenario2_temp} and \ref{fig:scenario2_control} show the temperature and control signal responses of the three controllers under the combined load ramp and valve leakage fault. This represents a realistic and challenging operating condition, combining both a predictable disturbance and an unmeasured fault.

The dashed red line represents the setpoint at approximately 515~$^\circ$C. The fixed-gain PI controller exhibits the poorest transient performance. It produces a large overshoot that peaks near 524~$^\circ$C (around +9~$^\circ$C above the setpoint, that would trigger safety alarms in practice), followed by a deep undershoot to about 510~$^\circ$C. These oscillatory excursions result in a long settling time and large transient error, even though the final value slowly drifts back toward the setpoint. Such behavior would be unacceptable in practice, as the overshoot and undershoot both exceed safe operating margins.  

The LSTM controller shows improved performance compared to the PI controller. Its overshoot is reduced, with a peak of about 521~$^\circ$C (around +6~$^\circ$C above the setpoint), and the oscillations are more damped, allowing it to recover more quickly. Nevertheless, because the leakage fault was not included in its training data, the LSTM cannot completely reject the disturbance. As a result, a small steady-state offset remains visible in the later part of the trajectory.  

The PINN controller delivers the best overall performance. Its physics-guided formulation enables real-time adaptation to the unmodeled fault. The overshoot is significantly reduced to less than 2~$^\circ$C, peaking at around 516–517~$^\circ$C, and the trajectory quickly damps to the setpoint with negligible steady-state error. Compared to the other controllers, the PINN achieves the fastest settling, the smallest overshoot, and the most accurate long-term tracking of the desired temperature.

The PI control input exhibits a stepwise profile that reflects the discrete adjustments of the physical actuator. It increases sharply during the transient around $t=1000$ to counteract the disturbance and then stabilizes in the range of 40--45 percent in steady state. The staircase behavior is typical of valve position changes and indicates relatively smooth actuation in practice.  

The LSTM-based control input introduces significant high-frequency fluctuations. While it rises to a similar magnitude during the transient and eventually stabilizes near the same operating point, the oscillatory nature of the signal is evident throughout. Such rapid variations could lead to unnecessary actuator stress and wear, even if the average performance remains comparable to the measured input.  

The PINN-based control input maintains smoother transitions than the LSTM. It reaches a comparable peak of about 75\% during the transient response and then settles near 40--45 in steady state, closely matching the actuator demand. Unlike the LSTM, the PINN avoids excessive oscillations, producing a stable and actuator-friendly input trajectory.

\subsubsection{Quantitative Performance Comparison}
The superiority of the adaptive controllers, especially the PINN, is quantified in Table \ref{tab:performance_metrics}. The KPIs are calculated for the duration of the combined scenario.

\begin{table}[h!]
	\centering
	\caption{Quantitative Performance Comparison of Controllers}
	\label{tab:performance_metrics}
	\begin{tabular}{lcccc}
		\hline
		\textbf{Controller} & \textbf{IAE [$^\circ$C$\cdot$s]} & \textbf{MO [$^\circ$C]} & \textbf{$T_s$ [s]} & \textbf{CEV $(kg/s)^2$} \\ \hline
		PI & 826.3 & 9 & 1855 & 1.36 \\
		LSTM & 685.7 & 6.2 & 1179 & 1.87 \\
		PINN & \textbf{314.8} & \textbf{1.83} & \textbf{805} & \textbf{0.82} \\ \hline
	\end{tabular}
\end{table}

The results in Table \ref{tab:performance_metrics} lead to the following conclusions:
\begin{itemize}
	\item Both adaptive controllers (LSTM and PINN) significantly outperform the conventional PI across all metrics.
	\item The PINN achieves a 54\% reduction in IAE compared to the LSTM, highlighting its ability to handle previously unseen faults due to its physics-informed design.
	\item The PINN consistently produces the smoothest control action (lowest CEV), which is critical for reducing actuator wear in industrial systems.
\end{itemize}

\begin{figure}[h!]
	\centering
	\includegraphics[width=0.95\linewidth]{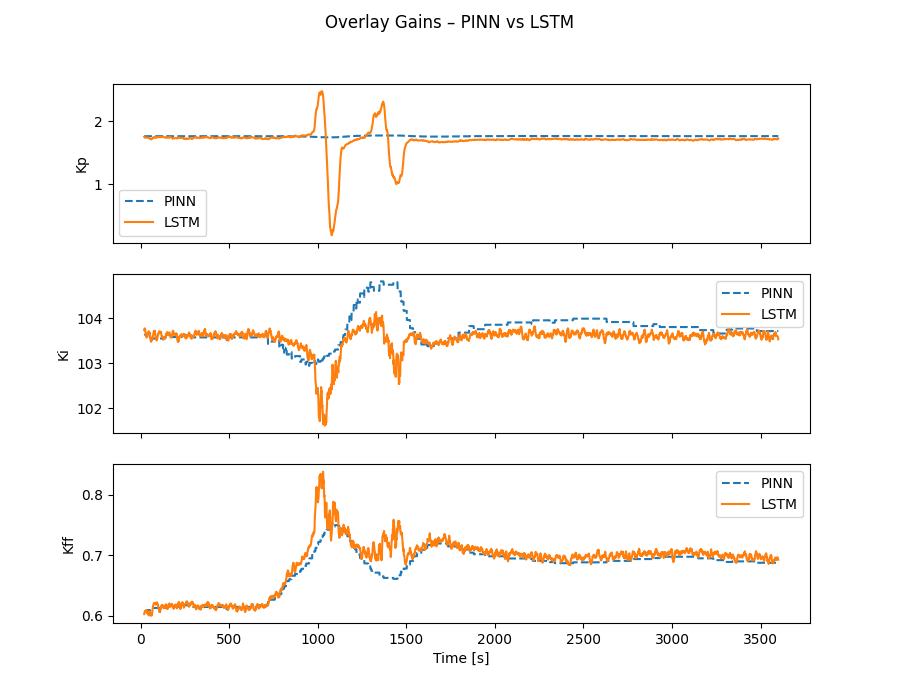}
	\caption{Evolution of the adaptive gains for the PINN and LSTM controllers during the combined load change and valve leakage fault scenario.}
	\label{fig:pinn_gains}
\end{figure}

The adaptive gains of both the PINN and LSTM controllers during the combined scenario are shown in Fig \ref{fig:pinn_gains}. For the PINN, the proportional gain $K_p$ slightly changes during the transient to improve responsiveness, the integral gain $K_i$ adapts to eliminate the steady-state error caused by the leakage, and the feedforward gain $K_{ff}$ adjusts to track the ramping exhaust temperature. The LSTM also adapts its gains in a less coordinated manner, highlighting the superior robustness and fault tolerance of the physics-informed approach.

\section{Conclusion}

This study conducted a comparative analysis of a data-driven LSTM network and a physics-informed neural network for adaptive temperature control of a HRSG superheater.

The results demonstrated that both advanced controllers significantly outperformed a conventional PI controller. The LSTM controller, trained on historical data, effectively handled load changes, exhibited degraded performance when faced with an unseen valve leakage fault, revealing a key limitation of purely data-driven approaches.

In contrast, the PINN controller, which integrates thermodynamic laws directly into its learning process, demonstrated superior robustness and fault tolerance. It successfully compensated for the unmodeled leakage in real-time, achieving a 54\% reduction in integral absolute error compared to the LSTM under fault conditions.

The key conclusion is that while LSTMs are powerful for modeling known operational patterns, PINNs provide a more robust and reliable framework for industrial control by ensuring physical consistency. This makes the physics-informed approach particularly valuable for maintaining safe and efficient operation in the presence of unforeseen faults and disturbances.

\bibliographystyle{ieeetr}
\bibliography{mainrefs}

\end{document}